\documentstyle[preprint,psfig,epsfig,lscape]{aastex}
\newcommand{\be}{\begin{equation}}
\newcommand{\ee}{\end{equation}}
\newcommand{\ba}{\begin{eqnarray}}
\newcommand{\ea}{\end{eqnarray}}

\newcommand{\nn}{\nonumber \\}

\newcommand{\n}{\mbox{\boldmath $n$}}

\def\gs{\mathrel{\raise1.16pt\hbox{$>$}\kern-7.0pt %
\lower3.06pt\hbox{{$\scriptstyle \sim$}}}}         %
\def\ls{\mathrel{\raise1.16pt\hbox{$<$}\kern-7.0pt %
\lower3.06pt\hbox{{$\scriptstyle \sim$}}}}         %

\shorttitle{GREAT10 Star Challenge}
\shortauthors{Kitching et al.}

\begin{document}

\title{Image Analysis for Cosmology :\\ Results from the GREAT10 Star Challenge}

\author{T. D. Kitching\altaffilmark{1}, B. Rowe\altaffilmark{2,3},
  M. Gill\altaffilmark{4,5,6}, C. Heymans\altaffilmark{1},
  R. Massey\altaffilmark{7}, D. Witherick\altaffilmark{2},
  F. Courbin\altaffilmark{8},
  K. Georgatzis\altaffilmark{9,10}, M. Gentile\altaffilmark{8},
  D. Gruen\altaffilmark{11,12},
  M. Kilbinger\altaffilmark{13,14}, G. L. Li\altaffilmark{15,16},
  A. P. Mariglis\altaffilmark{10}, G. Meylan\altaffilmark{8},
  A. Storkey\altaffilmark{10}, B. Xin\altaffilmark{16}}

\altaffiltext{1}{University of Edinburgh, Royal Observatory, Blackford Hill, Edinburgh, EH9 3HJ,
U.K. {tdk@roe.ac.uk}}
\altaffiltext{2}{Department of Physics and Astronomy, University College London, Gower Street, London, WC1E 6BT, U.K.}
\altaffiltext{3}{California Institute of Technology, 1200 E California Blvd, Pasadena CA 91125, USA}
\altaffiltext{4}{Center for Cosmology and AstroParticle Physics Physics Dept, The
Ohio State University, USA.}
\altaffiltext{5}{Kavli Institute for Particle Astrophysics \& Cosmology, Stanford, USA}
\altaffiltext{6}{Centro Brasileiro de Pesquisas FÃ­sicas,Rio de Janeiro, RJ, Brazil}
\altaffiltext{7}{Institute for Computational Cosmology, Durham University, South
Road, Durham, DH1 3LE, U.K.}
\altaffiltext{8}{Laboratoire d'Astrophysique, Ecole Polytechnique
  Federale de Lausanne (EPFL), Switzerland}
\altaffiltext{9}{Aalto University, Department of Information and Computer Science, P.O. Box 15400, FI-00076 Aalto, Finland}
\altaffiltext{10}{University of Edinburgh, School of Informatics, 10 Crichton Street, Edinburgh, EH8 9AB, United Kingdom}
\altaffiltext{11}{Department of Physics and Astronomy, 209 South 33rd Street,
University of Pennsylvania, Philadelphia, PA 19104, USA}
\altaffiltext{12}{University Observatory Munich, Scheinerstrasse 1, 81679 Munich, Germany}
\altaffiltext{13}{Excellence Cluster Universe, Boltzmannstr. 2, D-85748 Garching, Germany;
Universit\"ats-Sternwarte, Ludwig-Maximillians-Universit\"at
M\"unchen, Scheinerstr.~1, 81679 M\"unchen, Germany}
\altaffiltext{14}{CEA Saclay, Service d'Astrophysique (SAp), Orme des Merisiers, B\^at
709, F-91191 Gif-sur-Yvette, France}
\altaffiltext{15}{Purple Mountain Observatory, 2 West Beijing Road, Nanjing 210008, China}
\altaffiltext{16}{Department of Physics, Purdue University, 525 Northwestern Ave., West Lafayette, Indiana 47907, USA}

\begin{abstract}
We present the results from the first public blind PSF reconstruction
challenge, the GRavitational lEnsing Accuracy Testing 2010 (GREAT10)
Star Challenge. Reconstruction of a spatially varying PSF, sparsely
sampled by stars, at non-star positions is a critical part in the
image analysis for weak lensing where inaccuracies in the modelled
ellipticity $e$ and size $R^2$ can impact the ability to measure the shapes of
galaxies. This is of importance because weak lensing is a particularly 
sensitive probe of dark energy, and can be used to map the mass
distribution of large scale structure. Participants in the challenge were presented
with $27$,$500$ stars over $1300$ images subdivided into $26$
sets, where in each set a category change was made in the type or
spatial variation of the PSF. Thirty submissions were made by $9$ teams. 
The best methods reconstructed the PSF with an accuracy of $\sigma(e)\approx 2.5$x$10^{-4}$ and
$\sigma(R^2)/R^2\approx 7.4$x$10^{-4}$. For a fixed pixel scale narrower PSFs were found to be 
more difficult to model than larger PSFs, and the PSF
reconstruction was severely degraded with the inclusion of an atmospheric
turbulence model (although this result is likely to be a strong
function of the amplitude of the turbulence power spectrum). 
\end{abstract}

\keywords{Cosmology: observations, Methods: data analysis, Atmospheric
  effects, Techniques: image processing}

\section{Introduction}
\label{Introduction}
In this paper we present the results from the GRavitational lEnsing Accuracy Testing 2010 (GREAT10) Star Challenge.
GREAT10 was an image analysis challenge for cosmology that focused on
the task of measuring the shapes of distant galaxies. 
Light from distant galaxies is deflected during its journey to us via
gravitational lensing, and the images appear distorted into
characteristic patterns \citep{hu99,bs01}. 
The amount of distortion depends on the intervening distribution of
matter (including dark matter) and the geometry of spacetime (which is
currently governed by dark energy). 
Such measurements thus probe directly the invisible dark sector and
the fundamental nature of gravity --- see reviews by
\citet{arev,rrev,hrev,mrev,wrev}. 

All real imaging data are necessarily seen after convolution with
(i.e. blurring by) a telescope's Point Spread Function (PSF). 
The PSF arises from the finite aperture of the telescope, charge
diffusion within digital detectors, any imperfect elements along the
optical path, and turbulence in the Earth's atmosphere (unless the
telescope is in space). 
This increases the size of faint galaxies, and can spuriously change
their ellipticity by an order of magnitude more than gravitational
lensing \citep{bj02,henkpsf,sph08,sparsity,mhk12}. 
To recover the shape of the galaxy after only cosmological effects, it
is necessary to (1) model the PSF and (2) somehow correct for its
effect on the images of galaxies. 
The second half of this task has been widely addressed by teams
analysing individual surveys and, as a vital community effort, through
the public Shear TEsting Programme (STEP) \citep{step1,step2}, the 
GRavitational lEnsing Accuracy Testing (GREAT) galaxy challenges
\citep{great08,g10res} and the Mapping Dark Matter challenge \citep{kaggle}. 
The first task (modelling the PSF) has so far only been investigated
internally within teams
\citep[e.g.][]{bac03,hyg04,lvw05,rho07,sch10,henkpsf,lvw05,rowe10,jj05}. 
Here we present the results of the first blind, public trial of
methods to model and interpolate the PSF of a typical astronomical
telescope. 

The PSF in an astronomical image can be measured from stars that happen to fall inside the field of view.
Stars are so small that they are intrinsically point-like, and adopt the size and shape of the telescope's PSF.
However, the PSF typically varies across the field of view, and stars
only sparsely cover the extragalactic sky
\citep{jj05,jain06,cfhtpsf,lsstpsf}. 
It is therefore necessary to model the shapes of stars, then
interpolate their shapes to the locations of galaxies (where there is
necessarily not a bright star, because otherwise the galaxy could not
be seen). 
In practice the PSF also varies as a function of the wavelength of
observed light, due to diffraction, reflection and transmission effects 
in the telescope optics, filters and CCDs and so must also be
interpolated from the colours of the
stars to the colours of the galaxy \citep{cyp09,voi11,plaz}. Colour
dependence is an important second order effect but in this paper we do
not address this, focussing only on the primary changes in PSFs.

We simulated the spatial variation in the PSF of generic
but realistic ground-, balloon-, and space-based telescopes
\citep[][and see \url{www.greatchallenges.info}]{great10}. 
We realised a large suite of sparse stellar fields in these different
observing regimes, and publicly released {\em most} of the star
images. Entrants were asked to then reconstruct the images of the
missing stars on a pixel grid, at pre-defined locations. 
The performance of each entry was measured in real time using a single
number, `quality factor', which was designed to provide a crude ranking
such that it could not be reverse-engineered to reveal the full
solutions.
In this paper, we analyse in detail the quantitative performance
of $12$ distinct algorithms submitted to model and interpolate the
simulated PSFs. In particular we quantify how well the ellipticity and
size of a spatially varying PSF can be reconstructed in a blind
challenge.  

This paper is organised as follows. In Section~\ref{Description of the
  Simulations}, we describe the simulations and competition in
detail. In Section~\ref{Results}, we present results. We discuss and 
conclude in Section \ref{Conclusions}.

\section{Method}
\label{Description of the Simulations}
In this Section we describe the simulations and the competition. For a
full exposition of the background of the Star Challenge see \citet{great10}.

\subsection{Simulation Structure}
\label{Simulation Structure}
In the simulations we aimed to generate simplified representations of
possible observing scenarios and telescopes, such that through analysis we
could make general statements about how methods perform in a coarse-grained
sense in each of these categories.  

The simulations contained two possible types of PSF function: a Moffat
function (Moffat, 1969) and an Airy disk, parameterized by a FWHM size.
To simulate diffraction spikes caused by obscuration of the telescope
pupil the intensity distributions of these functions were optionally
combined with single-slit diffraction intensity patterns, 
approximating the effects of rectangular obscurations
in the pupil plane as would be caused by struts supporting a secondary
mirror.  The
dimensions of these single slit obscurations were chosen to produce
simulated PSFs of reasonable realism on visual inspection; for the
Airy disk this corresponded to a strut obscuration of width 4\% the
pupil diameter.  The configurations chosen for these diffraction spike
patterns were a `plus-sign' four-fold symmetric mask $+$, or an
`asterisk-sign' six-fold symmetric pattern $\ast$\footnote{We use 
the term `mask' to label such configurations, but we remind that
reader that for the $\ast$ pattern a telescope would only have $3$ struts
arranged in a trefoil shape - it is the slit diffraction that results in
six spikes in the images.}. The
combined pattern was then given a linear coordinate shear to create
elliptical PSF patterns, and the PSF spatial variation for any image
then contained three components, similar to the PSF described in
Kitching et al. (2012b: Appendix C, where we refer the reader to
Figures that show the PSF variation):
\begin{itemize} 
\item 
{\bf Static Component.} These were spatially constant across the image
and consisted of i) a Gaussian smoothing kernel that added to the PSF
size, this had a variance of $0.1$ present in all images, ii) a static
additive ellipticity component of $0.05$ in $e_{1,{\rm PSF}}$ and
$e_{2,{\rm PSF}}$, to simulate tracking error ($e_1$ and $e_2$ are
defined in Section \ref{Competition Structure}). Details are explained
in Kitching et al. (2012b). 
\item 
{\bf Deterministic Component.} This was to simulate the impact of the
telescope on the spatially varying PSF size and ellipticity. We used
the \citet{jsj08} model with fiducial parameters given in \citet{g10res}
$(a_0= 0.014$, $a_1= 0.0005$, $d_0=-0.006$, $d_1= 0.001$,
$c_0=-0.010)$, which is dominated by primary astigmatism ($a_0$), 
primary de-focus ($d_0$) and coma ($c_0$).
\item 
{\bf Random Component.} To simulate the random turbulent effect of the
atmosphere we additionally included a random
Gaussian field in some images in the ellipticity only, with a Kolmogorov-like power spectrum of 
$C_{\ell}=\ell^{-11/6}$. In fact subsquent to the formulation
of this challenge, and launch in 2010, Heymans et al. (2012) found that $C_{\ell}\propto
\ell^{-11/3}$, the exact power was not know
accurately beforehand hence we refer to the $C_{\ell}\propto
\ell^{-11/6}$ as Kolomogorov-like; this is approximately similar to short
exposures from a ground-based observatory for a Moffat PSF, or balloon-based if an Airy
PSF is used. We note that the amplitude of the power is also very high, corresponding to
exposures of $\simeq 1$ second (see Heymans et al., 2012): we leave an investigation into the
impact of varying amplitudes of Kolmogorov power to future work. 
\end{itemize}
The integration of the PSF
intensity distribution onto square pixels was achieved by
multiplication with a Sinc function in Fourier space (equivalent to
convolution with a square boxcar function in real space), followed by
sampling at the locations of pixel centres. 

\subsection{Data Structure}
The simulation was designed within the constraint that both the
download size of the simulation and the upload size of the submissions
should be manageable (we limited the download size to $50$ Gb). 
Participants were provided with FITS \citep{fits} images containing 
`known-stars' that were delta functions convolved with a spatially varying
PSF. Each star within each image was embedded in a postage stamp of
48x48 pixels, and to reduce the size of the images there was no noise
in between postage stamps. 
Participants were then asked to submit a 2D image of the
reconstructed PSF at positions in between the known-stars; these
positions were provided as a catalogue of `asked-star' positions. 
Participants were asked to submit FITS
cubes of the reconstructed PSFs (the x and y dimensions
representing the 2D image and the z dimension varying the asked-star
positions).

For each image $1000$ asked-stars were required. 
The images were subdivided into $26$ sets of $50$ images where in each set
the the properties of the spatial variation, telescope and static
components of the PSF were kept statistically constant, 
but each had a different realisation of any random component, and
each also had the asked- and known-star positions varying. The
properties of each set are summarised in Table \ref{sets}. One aspect to note is that when varying the
size of the PSF in the total flux was kept constant for each profile;
with an integrated signal to noise of $100$. 

\clearpage
\begin{table}
\begin{center}
\caption{The properties of each set of images. Details are described
  in Section \ref{Simulation Structure}. Each category allows a
  different test: PSF Size allows us to test
  under-sampling; Atmosphere tests ground-based exposure time dependence;
  $N_{\rm Stars}$ tests spatial sampling; Mask tests telescope
  structure dependence; PSF-Type tests the impact of high spatial
  frequencies in the PSF profile vs smooth profiles; Telescope
  variation allows us to test the impact of three typical distortions
  found in data. The set order was semi-random so as to prevent
  participants exploiting any pattern in the set numbering. We label
  the fiducial sets for the Moffat and Airy profiles. \label{sets}}
\begin{tabular}{crrrrrr}
\tableline\tableline
Set & Atmosphere & PSF-Type & Mask & $N_{\rm Stars}$ & PSF
Size/pixels&Telescope Variation\\
\tableline
1 (fid. Airy) & No  & Airy & None & 1000 & 3& None \\
2  & No  & Airy &  $+$ & 1000 & 3& None \\
3  & No  & Airy &$\ast$& 1000 & 3& None \\
4  & No  & Airy & None & 2000 & 3& None \\
5  & No  & Airy & None & 500  & 3& None \\
6  & No  & Airy & None & 1000 & 1.5& None \\
7  & No  & Airy & None & 1000 & 6& None \\
8 (fid. Moffat) & No  & Moffat & None & 1000 & 3& None \\
9  & Yes & Airy & None & 1000 & 3& None \\
10 & Yes & Moffat      &  $+$ & 1000 & 3& None \\
11 & Yes & Moffat      &$\ast$& 1000 & 3& None \\
12 & Yes & Moffat      & None & 2000 & 3& None \\
13 & Yes & Moffat      & None & 500  & 3& None \\
14 & Yes & Moffat      & None & 1000 & 1.5& None \\
15 & Yes & Moffat      & None & 1000 & 6& None \\
16 & No  & Airy & None & 1000 & 3&astigmatism $a_0$ \\
17 & Yes & Moffat      & None & 1000 & 3&astigmatism $a_0$ \\
18 & No  & Airy & None & 1000 & 3&de-focus $d_0$ \\
19 & Yes & Moffat      & None & 1000 & 3&de-focus $d_0$ \\
20 & No  & Airy & None & 1000 & 3&coma $c_0$ \\
21 & Yes & Moffat      & None & 1000 & 3&coma $c_0$ \\
22 & No  & Moffat      & $+$  & 1000 & 3& None \\
23 & No  & Moffat      &$\ast$& 1000 & 3& None \\
24 & No  & Moffat      & None & 2000 & 3& None \\
25 & No  & Moffat      & None & 500  & 3& None \\
26 & No  & Moffat      & None & 1000 & 1.5& None \\
\tableline
\end{tabular}
\end{center}
\end{table}

\subsection{Competition Structure}
\label{Competition Structure}
The competition started in December 2010 and ran for 9 months until
September 2011; this was concurrent with the GREAT10 Galaxy challenge
\citep{great10,g10res}. As stated previously 
the total simulation size was $\sim 50$ Gb and the
total size of the uploaded submissions was $\sim 1$ Gb (we allowed
participants to tar, zip or FITS-compress\footnote{{\tt
    http://heasarc.nasa.gov/fitsio/fpack/}} submissions to
reduce size). Data and example code were provided online for
participants\footnote{{\tt http://great.roe.ac.uk/data}}.

The two parameters of the PSF that most directly impact the ability
to interpret observations of galaxies are the ellipticity and the size
of the PSF; any residual difference between the ellipticity or size of
true PSF, and the respective quantities of the modelled PSF at any
particular position, will result in errors and biases in parameters
assigned to any galaxy at that position. 
Weak gravitational lensing is particularly sensitive to
these types of error \citep{mhk12,sph08,sparsity}. 
The ellipticity and size are defined here using the second order brightness moments of the image as 
\be 
q_{ij}=\frac{\sum_p w_p I_p (\theta_i - \bar \theta_i)(\theta_j - \bar
  \theta_j)}{\sum_p w_p I_p}, \,\,\,\, i,j\in\{1,2\},
\ee
where the sums are over pixels, $I_p$ is the flux in the $p^{\rm th}$
pixel and $\theta$ is a pixel position ($\theta_1=x_p$ and
$\theta_2=y_p$). 
In order to regularise the results with regard to the impact of noise
but not to constrain the interpretation to compact objects in the
postage stamp, we include a weight function $w_p$ chosen to be a broad
Gaussian with a width of $24$ pixels (we leave an investigation of
how results vary as a function of weight for future work). These are
almost unweighted quadrupole moments in this respect, and as a result, 
smooth analytical functions may be favoured compared to models that try to reproduce 
details in the wings of the PSF. The weighted ellipticity (or
technically the `polarisability') for a PSF in complex notation is defined as 
\ba
e &=& \frac{q_{11} - q_{22} + 2{\rm i}q_{12}}{q_{11} + q_{22} + 2(q_{11}q_{22}-q^2_{12})^{1/2}}\nn
\ea
where we have used a definition of ellipticity $|e|=(1-r)(1+r)^{-1}$,
where $r$ is the ratio of minor to major axes of the ellipse. 
For the weighted size we have a similar expression 
\begin{equation}
R^{2} = q_{11} + q_{22}.
\end{equation}
We can calculate the variance between the ellipticity of the model
and true PSF $\sigma^2(e)\equiv \langle(e - e_{\rm
  PSF}^{{\rm t}})^2\rangle$ and similarly for the size
$\sigma^2(R)\equiv\langle(R - R_{\rm
  PSF}^{{\rm t}})^2\rangle$. Submissions were scored in real-time on a leaderboard that displayed
the metric $P\equiv \frac{1}{\frac{1}{2}\langle\sigma^2(R)+\sigma^2(e)\rangle}$
where the average was taken over images in a set but not over
objects asked-star positions, such
that a mean variance of $10^{-3}$ in both ellipticity and size would
have $P\sim 1.0$. 

The $P$ metric, whilst indicatively ranking the methods, does not offer
any insight into the performance of a method on ellipticity and size
reconstruction. In this paper we will present quantities that relate
to the principal properties of the PSF more directly. These are the
standard deviation of mean of the residuals of the ellipticity $\sigma(e)$ and size-squared 
$\sigma(R^2)/R^2$ over all asked-stars i.e. we compute the error on the mean of the
residuals (the sample
variance computed using centred second order moments). We assume that
any mean bias could be removed through cross-validation, in this sense it
is a generous analysis to those methods with a mean
residual. We average these quantities
over the $50$ images in each set, but in fact for all methods we find
that the fractional error between images in a set is $\ls 10\%$.  

\section{Results}
\label{Results}
\begin{figure*}
\epsscale{.90}
\plotone{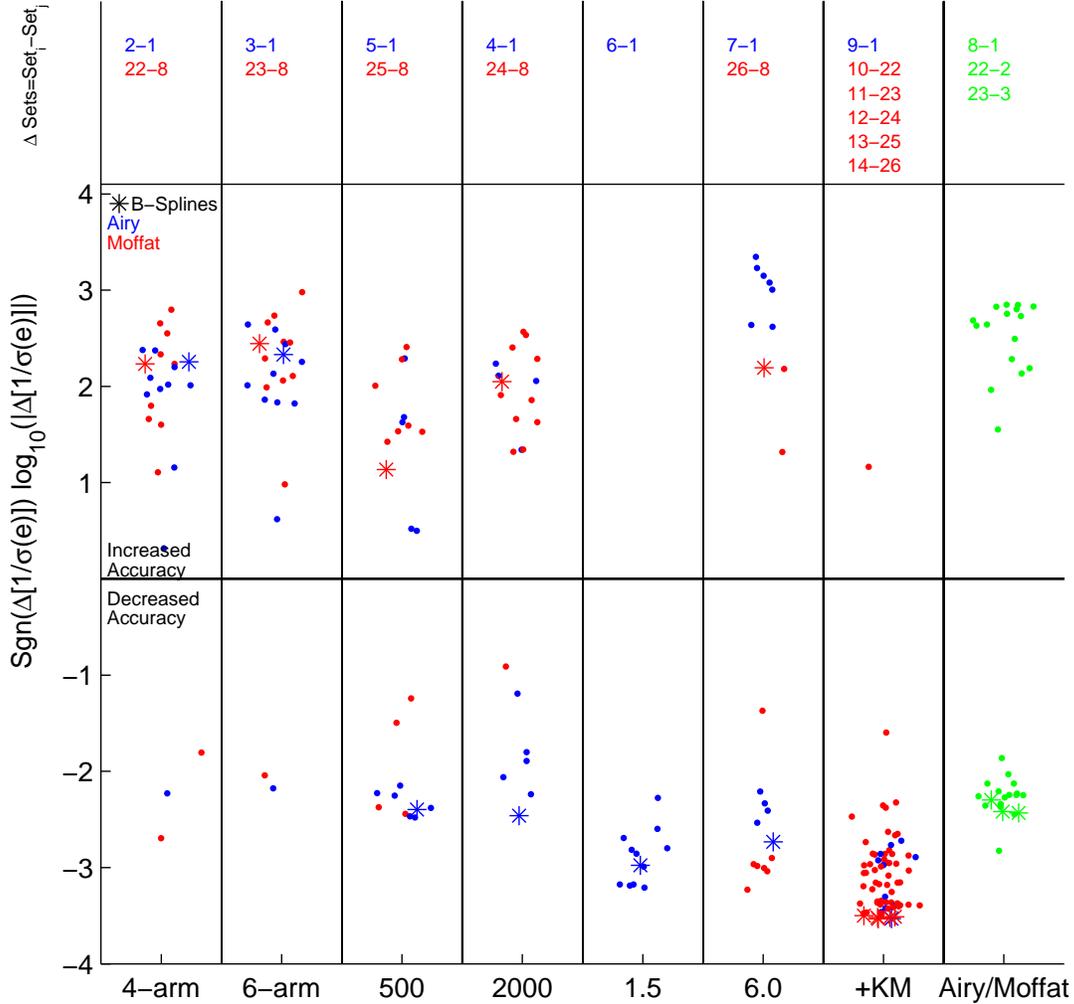}
\caption{The change in the inverse variance in the residual ellipticity for
  each method for each category varied.  The sets used in differencing
  the categories are shown in the upper panels (Set$_i$-Set$_j$), and
  we refer the reader to Table \ref{sets}. Each point represents a
  method, the stars represent method B-Spline, points
  within a bin are randomised within an x-bin for clarity. The log of the
  change is shown, with the sign preserved
  (i.e. $sgn[x]\log_{10}[|x|]$ where $x=(1/\sigma(e)_{\rm
    fiducial})-(1/\sigma(e))$) so that negative values
  represent a decrease in accuracy and positive values an increase in
  accuracy. The first seven vertical panels show changes for the
  Moffat (red) and Airy profile (blue), the rightmost panel
  shows the change in accuracy when the profile is changed from Airy
  to Moffat but all other aspects of the PSF at kept the same. The parameters varied are the mask (4-arm or 6-arm;
  changed from no mask),
  number of stars (500 or 2000; changed from 1000), PSF size (1.5 or
  6.0 pixels; changed from 3.0 pixels) and the addition of 
  Kolmogorov power in ellipticity. \label{fig00}}
\end{figure*}

\begin{figure*}
\epsscale{.90}
\plotone{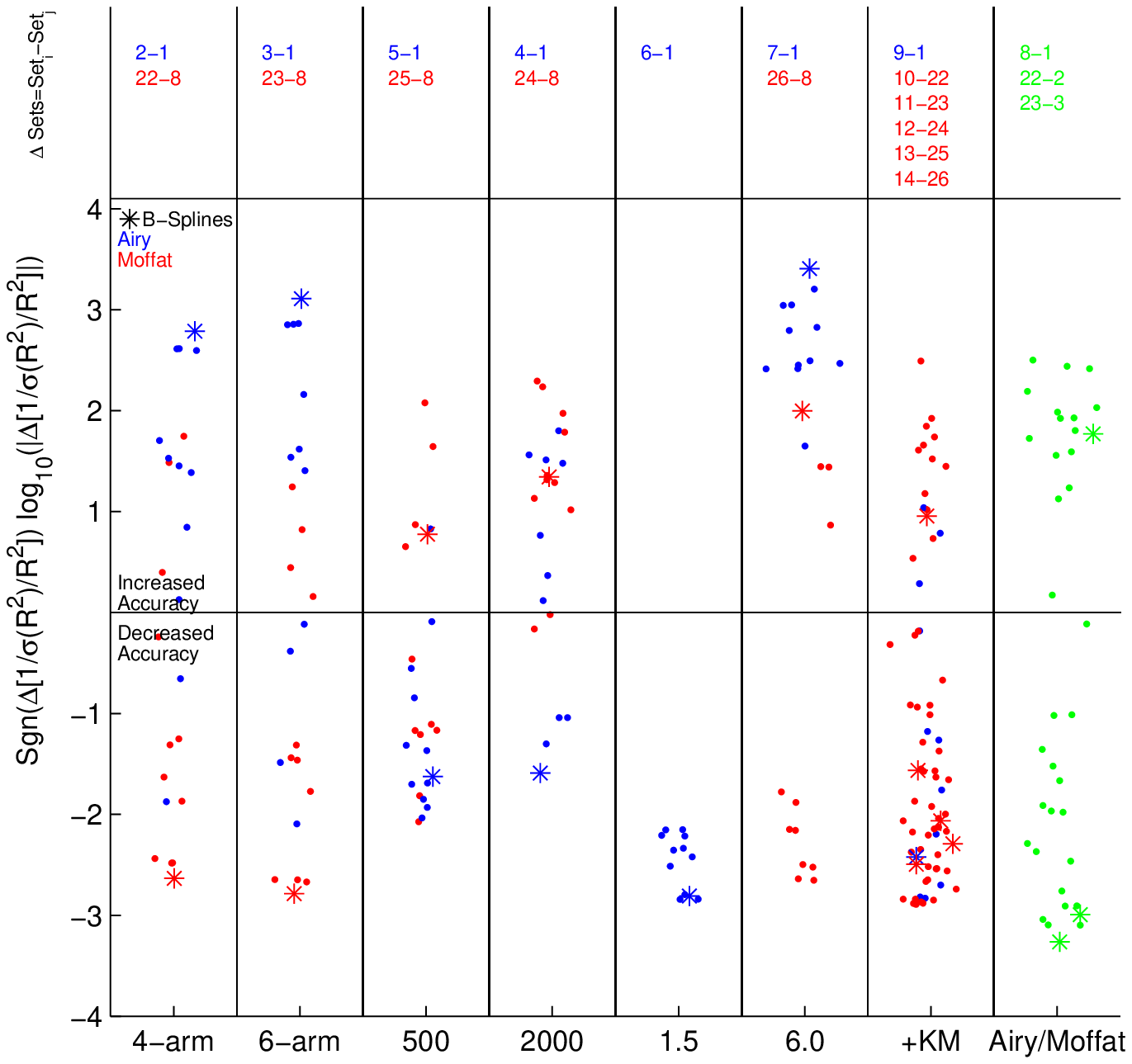}
\caption{The change in the inverse variance in the residual size-squared for
  each method for each category varied. The sets used in differencing
  the categories are shown in the upper panels (Set$_i$-Set$_j$), and
  we refer the reader to Table \ref{sets}. Each point represents a
  method, the stars represent method B-Spline, points
  within a bin are randomised in within an x-bin for clarity. The log of the
  change is shown, with the sign preserved
  (i.e. $sgn[x]\log_{10}[|x|]$ where $x=(1/[\sigma(R^2)/R^2]_{\rm
    fiducial})-(1/[\sigma(R^2)/R^2])$) so that negative values
  represent a decrease in accuracy and positive values an increase in
  accuracy. The first seven vertical panels show changes for the
  Moffat (red) and Airy profile (blue), the rightmost panel
  shows the change in accuracy when the profile is changed from Airy
  to Moffat but all other aspects of the PSF at kept the same. The parameters varied are the mask (4-arm or 6-arm;
  changed from no mask),
  number of stars (500 or 2000; changed from 1000), PSF size (1.5 or
  6.0 pixels; changed from 3.0 pixels) and the addition of 
  Kolmogorov power in ellipticity. \label{fig0}}
\end{figure*}

\begin{figure*}
\epsscale{.48}
\plotone{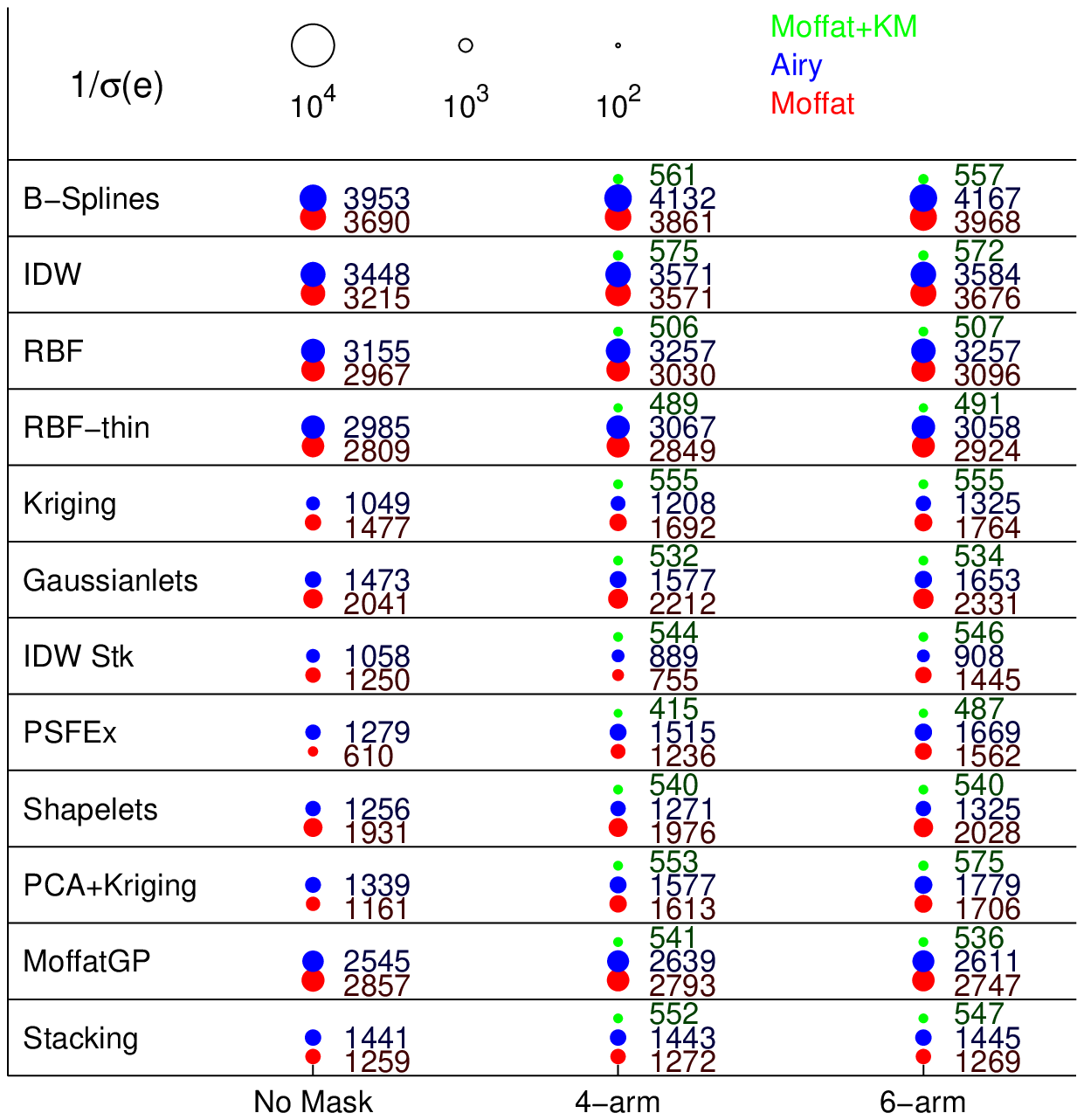}
\epsscale{.48}
\plotone{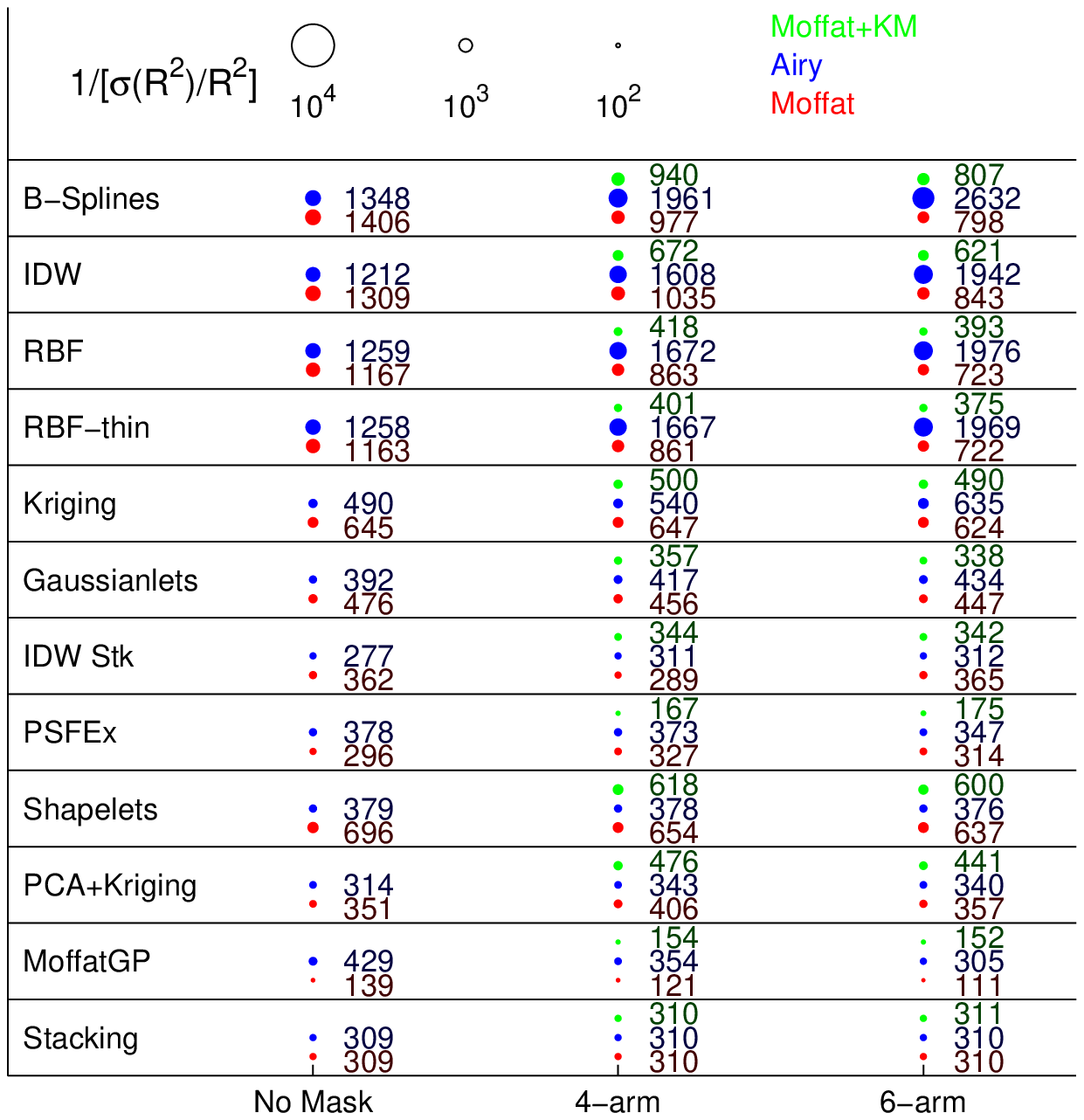}
\caption{The inverse variance in the residual ellipticity and size-squared for
  each method (horizontal panels) for the three mask cases 
  (no mask, 4-arm $+$ and 6-arm
  $\ast$) for the Moffat-plus-Kolmogorov case (green), Moffat with no
  Kolmogorov (red), and the Airy (blue) profile. The circles
  represent the inverse variance of the residual ellipticity and
  size-squared where the area scales in proportion to these
  parameters and the numbers are given next to each circle; 
  a key is given in the top panel. Where no number/circle is provided
  there was no set for this combination of PSF type and mask
  type. Fractional errors on the inverse variances are $\approx 10\%$
  for all methods.\label{fig1}}
\end{figure*}

\begin{figure*}
\epsscale{.48}
\plotone{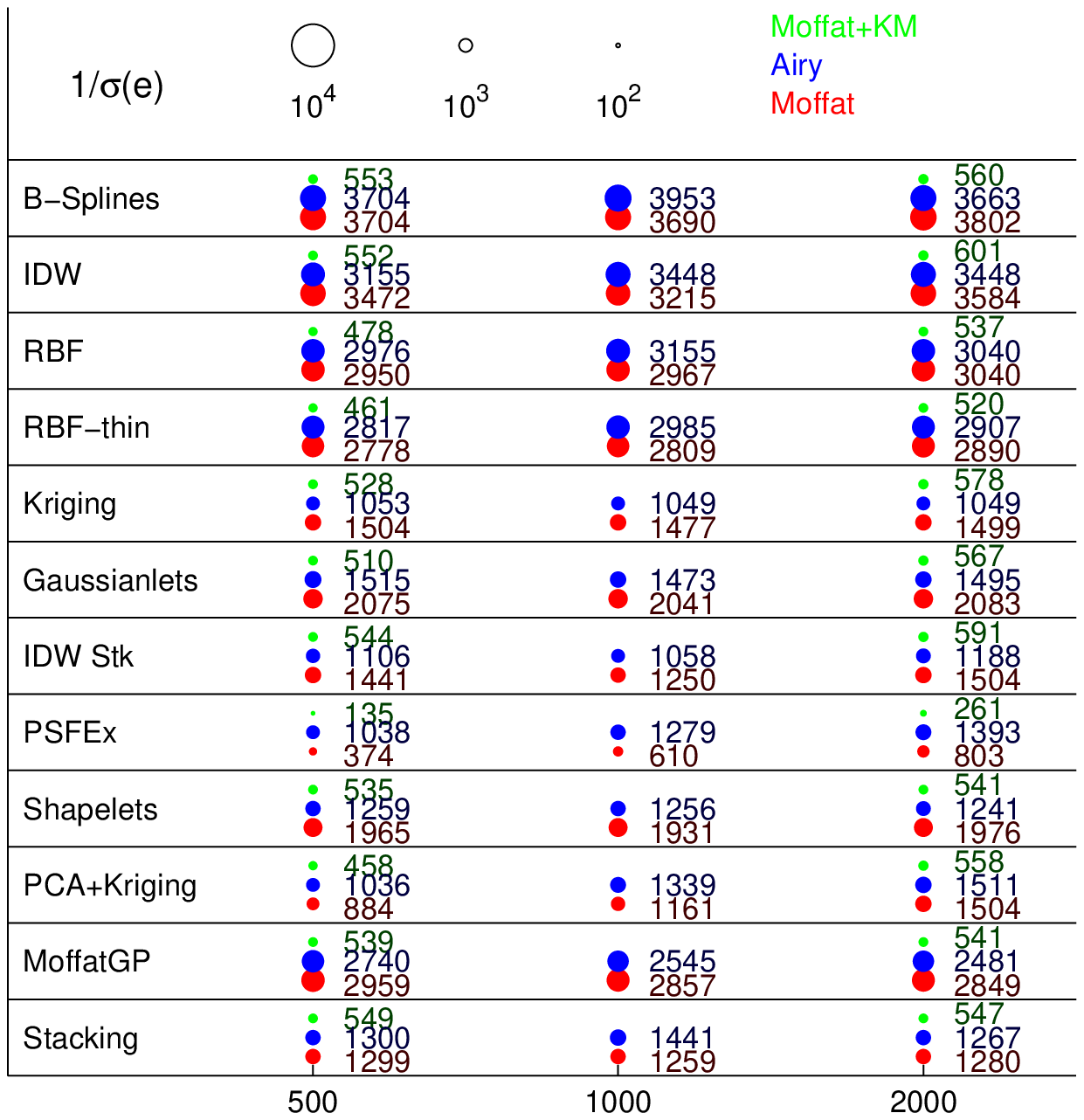}
\epsscale{.48}
\plotone{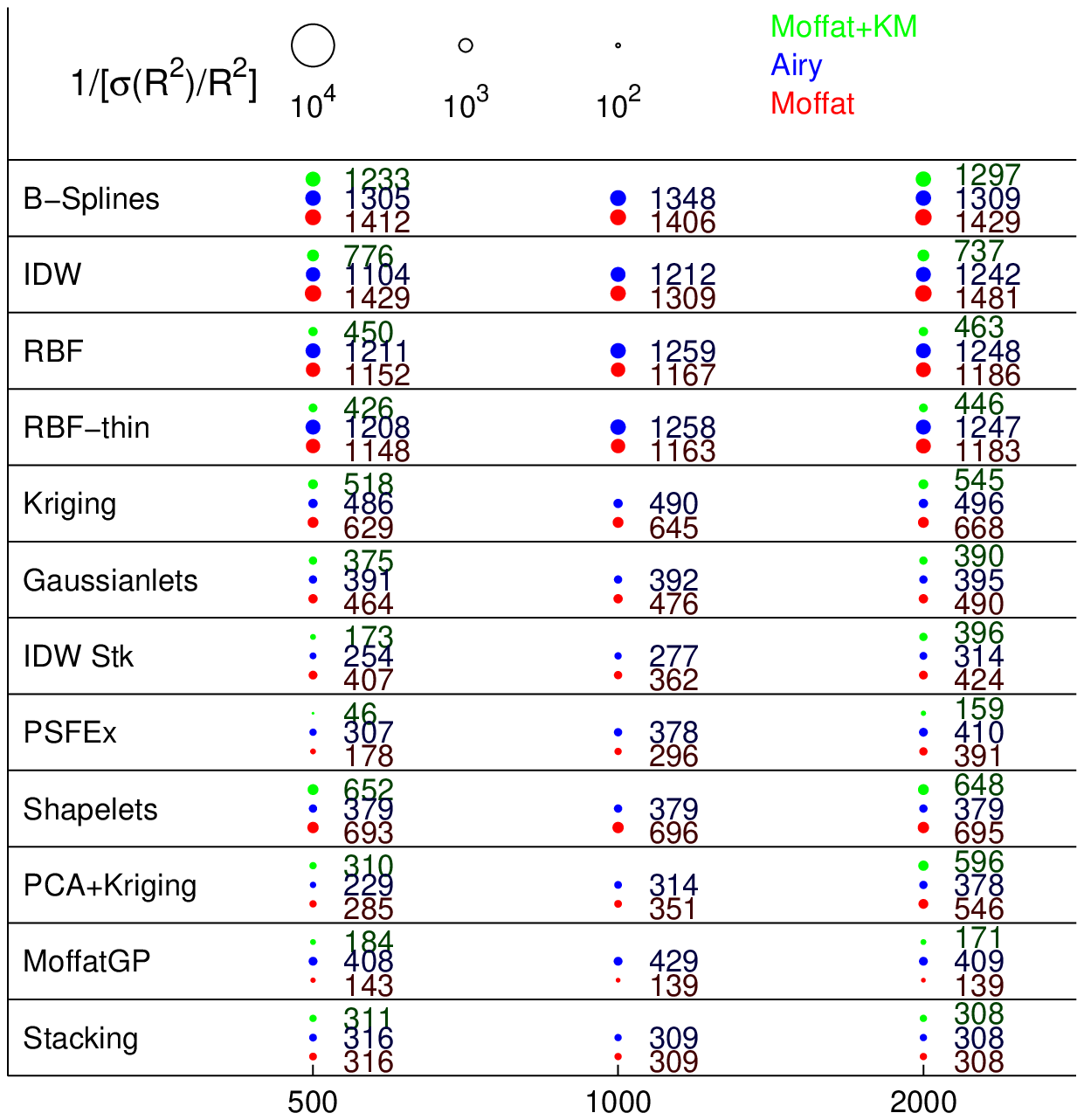}
\caption{The inverse variance in the residual ellipticity and size-squared for
  each method (horizontal panels) for the three known-star number cases 
  (500, 1000, 2000 stars) for the Moffat-plus-Kolmogorov case (green), Moffat with no
  Kolmogorov (red), and the Airy (blue) profile. The circles
  represent the inverse variance of the residual ellipticity and
  size-squared where the area scales in proportion to these
  parameters and the numbers are given next to each circle; 
  a key is given in the top panel. Where no number/circle is provided
  there was no set for this combination of PSF type and number of
  stars. 
  Fractional errors on the inverse variances are $\approx 10\%$
  for all methods.\label{fig2}} 
\end{figure*}

\begin{figure*}
\epsscale{.48}
\plotone{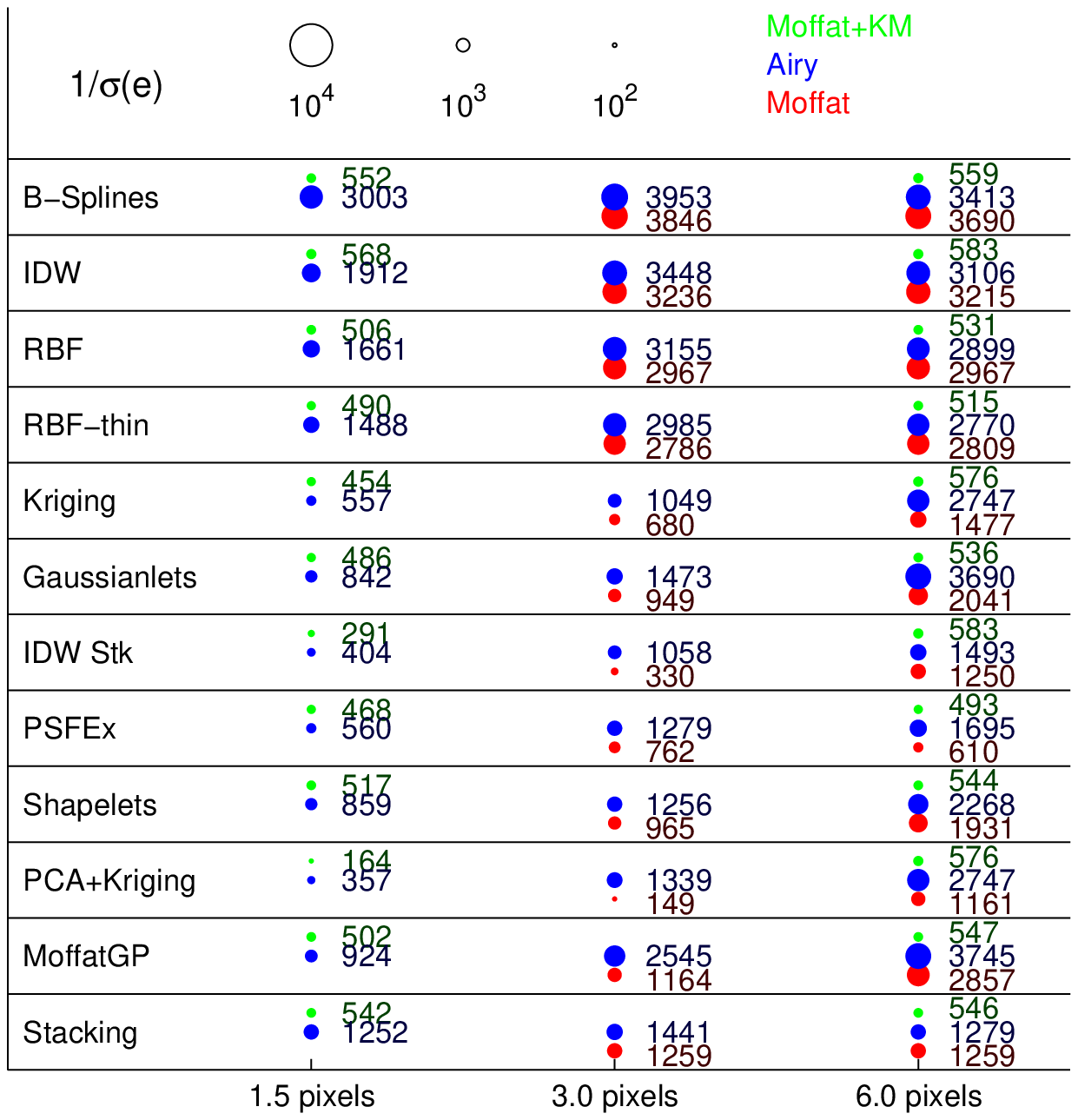}
\epsscale{.48}
\plotone{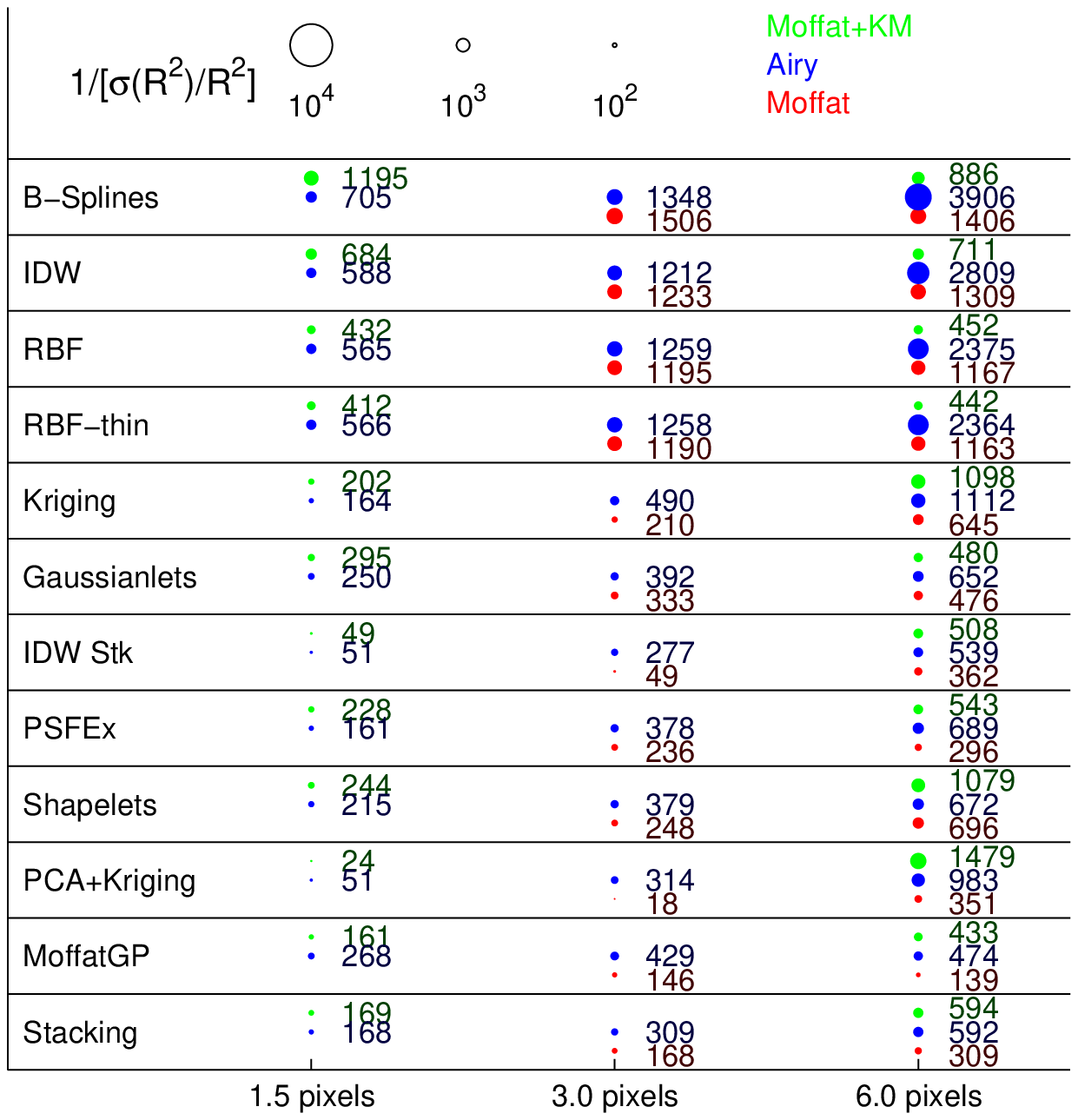}
\caption{The inverse variance in the residual ellipticity and size-squared for
  each method (horizontal panels) for the three PSF size cases 
  (1.5, 3.0 and 6.0 pixels) for the Moffat-plus-Kolmogorov case (green), Moffat with no
  Kolmogorov (red), and the Airy (blue) profile. The circles
  represent the inverse variance of the residual ellipticity and
  size-squared where the area scales in proportion to these
  parameters and the numbers are given next to each circle; 
  a key is given in the top panel. Where no number/circle is provided
  there was no set for this combination of PSF type and PSF
  size. Fractional errors on the inverse variances are $\approx 10\%$
  for all methods. \label{fig3}}
\end{figure*}

\begin{figure*}
\epsscale{.48}
\plotone{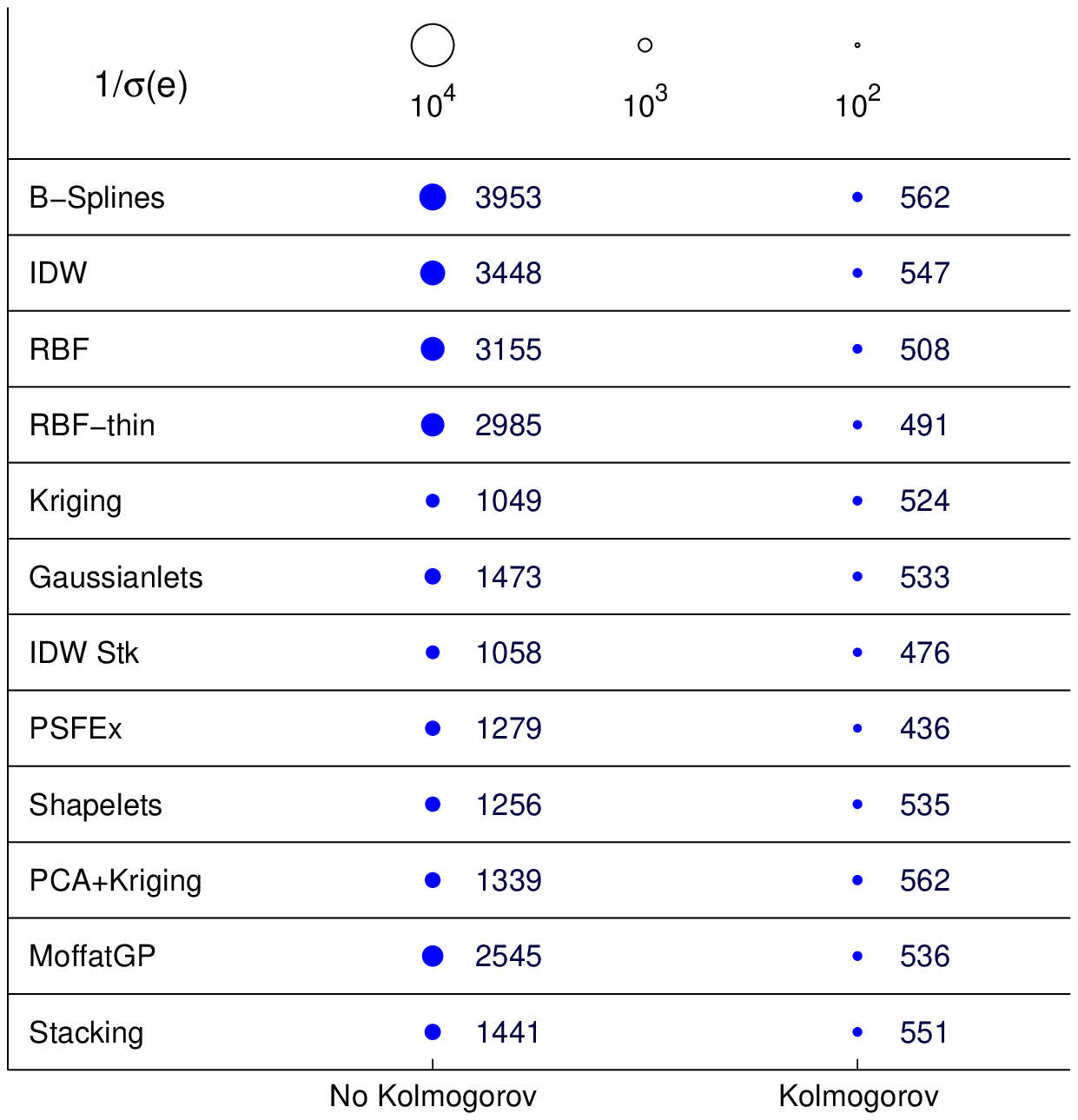}
\epsscale{.48}
\plotone{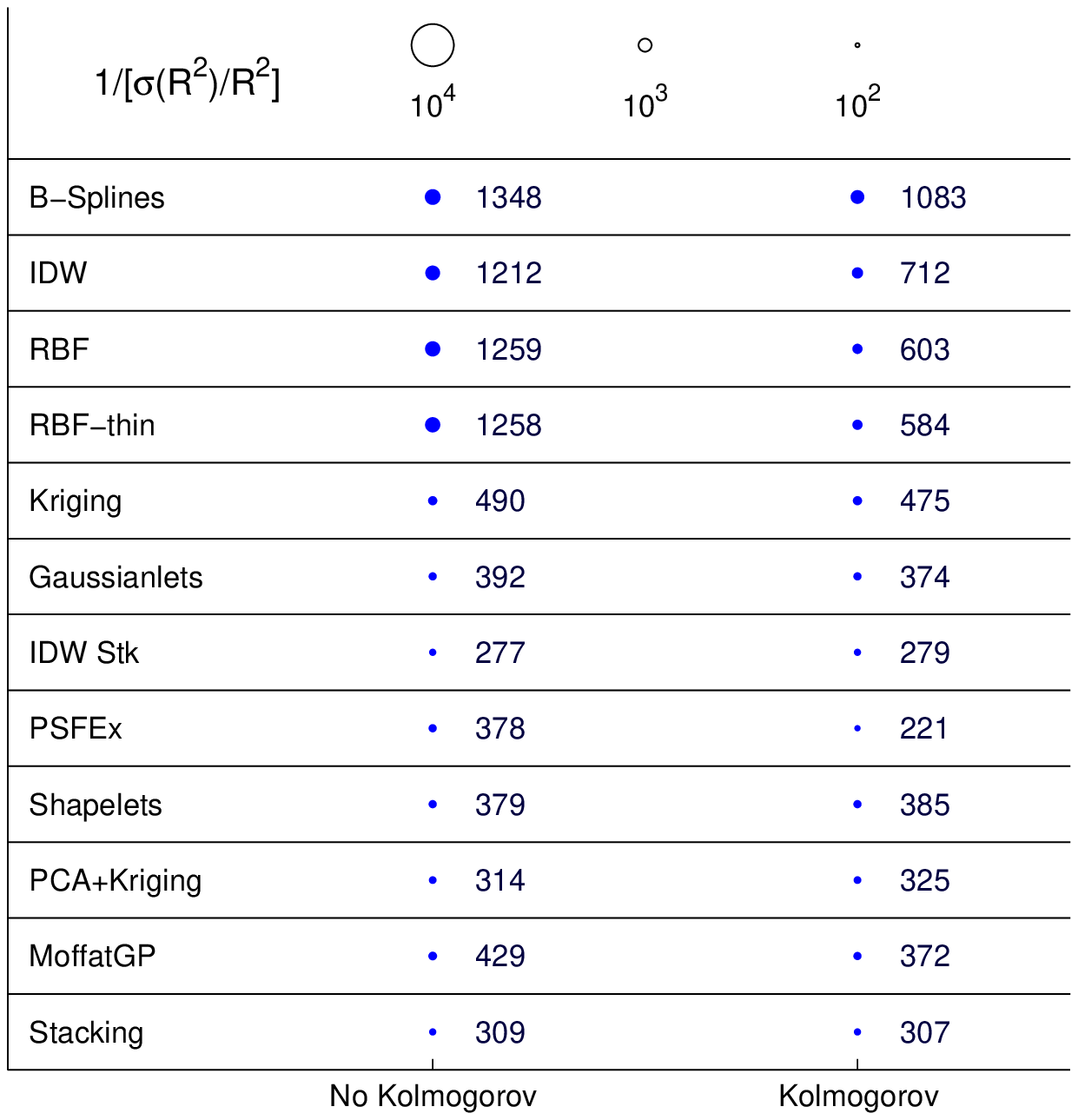}
\caption{The inverse variance in the residual ellipticity and size-squared for
  each method (horizontal panels) for the two cases 
  where a Kolmogorov power is added, or not to a set with a Airy
  (blue) profile. The circles
  represent the inverse variance of the residual ellipticity and
  size-squared where the area scales in proportion to these
  parameters and the numbers are given next to each circle; 
  a key is given in the top panel. Fractional errors on the inverse variances are $\approx 10\%$
  for all methods.\label{fig4}}
\end{figure*}

\begin{figure*}
\epsscale{.48}
\plotone{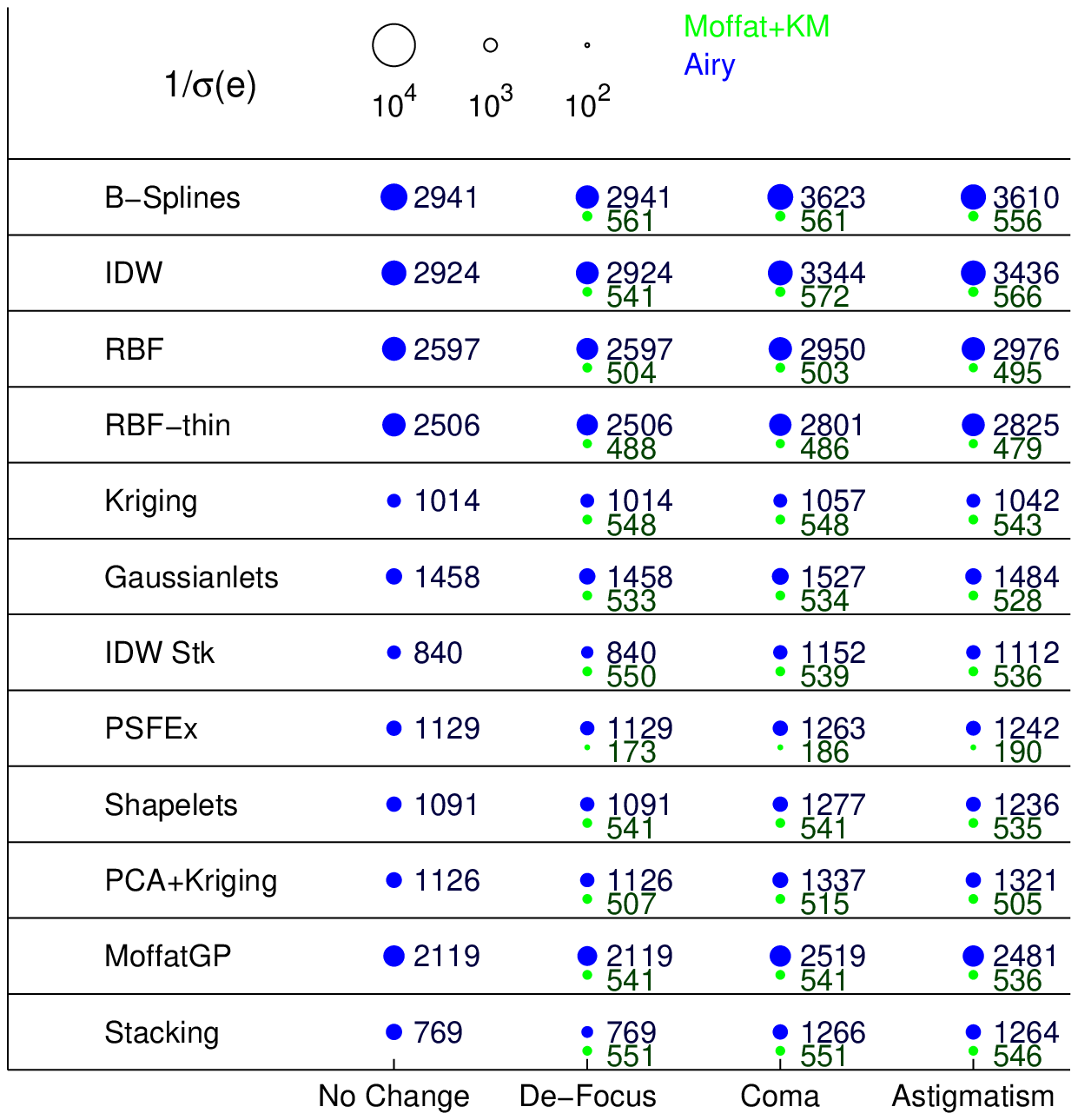}
\epsscale{.48}
\plotone{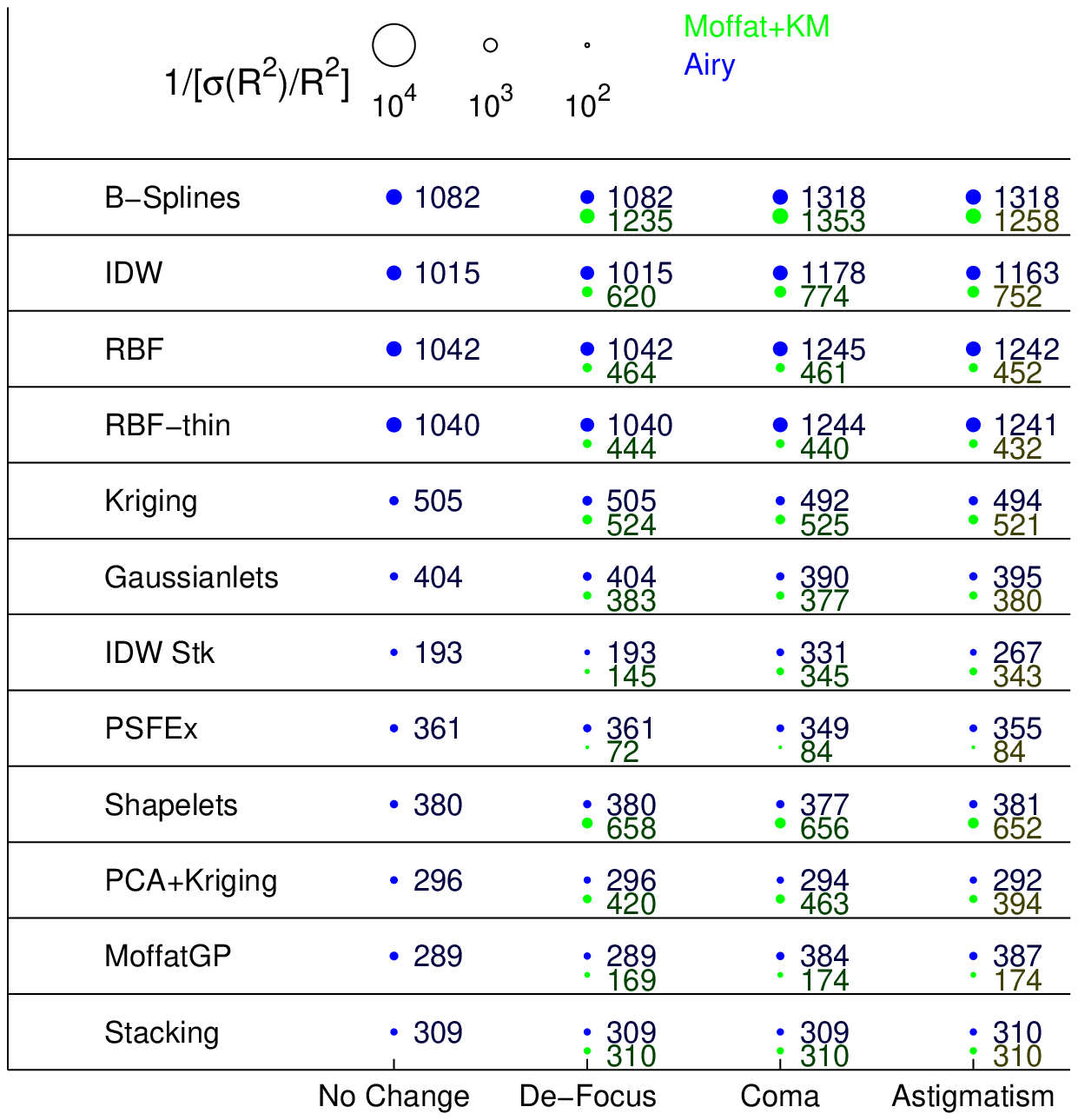}
\caption{The inverse variance in the residual ellipticity and size-squared for
  each method (horizontal panels) for the cases 
  where the telescope parameters are varied for the Airy
  (blue) profile and the Moffat-plus-Kolmogorov profile (green). The circles
  represent the inverse variance of the residual ellipticity and
  size-squared where the area scales in proportion to these
  parameters and the numbers are given next to each circle; 
  a key is given in the top panel.  Where no number/circle is provided
  there was no set for this combination of PSF type and telescope
  parameter. Fractional errors on the inverse variances are $\approx 10\%$
  for all methods.\label{fig5}}
\end{figure*}

In total $30$ submissions were made from $9$ teams. As a baseline
benchmark, a method in which all stars were simply stacked in an image
was created, where no spatial variation in the reconstructed stars was
present. Several methods generated low scores due to misunderstanding of
simulation details, resulting in scores below the benchmark, 
and in this paper we summarise only those not affected by these issues. 
In the following we choose the best performing submission, for size, for each of the
$12$ distinct method entries. All of the submitted methods are
described in Appendix A. We show the results on the fiducial Airy set (set
1 in Table \ref{sets}) and the fiducial Moffat set (set 8 in Table
\ref{sets}) in Tables \ref{set1} and \ref{set8} respectively. In
Figures \ref{fig00} and \ref{fig0} we present general behaviours of
methods over the sets as categories were change, but for a
quantitative presentation of each method we refer the reader to
Figures \ref{fig1} to
\ref{fig5} where we show pictographic tables of results. 

\clearpage 
\begin{table}
\begin{center}
\caption{The results for ellipticity and size-squared on set 1 (the
  fiducial Airy set) for each method tested in this paper.\label{set1}}
\begin{tabular}{crrrrrr}
\tableline\tableline
Method Name & $1/\sigma(e)$ & $\sigma(e)/10^{-4}$& $1/[\sigma(R^2)/R^2]$ & $[\sigma(R^2)/R^2]/10^{-3}$\\
\tableline
B-Splines          & $ 3953 $ & $ 2.53 $ & $ 1348 $ & $ 0.742 $\\ 
IDW                & $ 3448 $ & $ 2.90 $ & $ 1212 $ & $ 0.825 $\\ 
RBF                & $ 3155 $ & $ 3.17 $ & $ 1259 $ & $ 0.794 $\\ 
RBF-thin           & $ 2985 $ & $ 3.35 $ & $ 1258 $ & $ 0.795 $\\ 
Kriging            & $ 1049 $ & $ 9.53 $ & $ 490 $ & $ 2.042 $\\ 
Gaussianlets       & $ 1473 $ & $ 6.79 $ & $ 392 $ & $ 2.548 $\\ 
IDW Stk            & $ 1058 $ & $ 9.45 $ & $ 277 $ & $ 3.604 $\\ 
PSFEx              & $ 1279 $ & $ 7.82 $ & $ 378 $ & $ 2.647 $\\ 
Shapelets          & $ 1256 $ & $ 7.96 $ & $ 379 $ & $ 2.642 $\\ 
PCA+Kriging        & $ 1339 $ & $ 7.47 $ & $ 314 $ & $ 3.180 $\\ 
MoffatGP           & $ 2545 $ & $ 3.93 $ & $ 429 $ & $ 2.331 $\\ 
Stacking           & $ 1441 $ & $ 6.94 $ & $ 309 $ & $ 3.237 $\\ 
\tableline
\end{tabular}
\end{center}
\end{table}
\begin{table}
\begin{center}
\caption{The results for ellipticity and size-squared on set 8 (the
  fiducial Moffat set) for each method tested in this paper.\label{set8}}
\begin{tabular}{crrrrrr}
\tableline\tableline
Method Name & $1/\sigma(e)$ & $\sigma(e)/10^{-4}$& $1/[\sigma(R^2)/R^2]$ & $[\sigma(R^2)/R^2]/10^{-3}$\\
\tableline
B-Splines          & $ 3690 $ & $ 2.71 $ & $ 1406 $ & $ 0.711 $\\ 
IDW                & $ 3215 $ & $ 3.11 $ & $ 1309 $ & $ 0.764 $\\ 
RBF                & $ 2967 $ & $ 3.37 $ & $ 1167 $ & $ 0.857 $\\ 
RBF-thin           & $ 2809 $ & $ 3.56 $ & $ 1163 $ & $ 0.860 $\\ 
Kriging            & $ 1477 $ & $ 6.77 $ & $ 645 $ & $ 1.551 $\\ 
Gaussianlets       & $ 2041 $ & $ 4.90 $ & $ 476 $ & $ 2.099 $\\ 
IDW Stk            & $ 1250 $ & $ 8 $ & $ 362 $ & $ 2.759 $\\ 
PSFEx              & $ 610 $ & $ 16.40 $ & $ 296 $ & $ 3.374 $\\ 
Shapelets          & $ 1931 $ & $ 5.18 $ & $ 696 $ & $ 1.436 $\\ 
PCA+Kriging        & $ 1161 $ & $ 8.61 $ & $ 351 $ & $ 2.853 $\\ 
MoffatGP           & $ 2857 $ & $ 3.50 $ & $ 139 $ & $ 7.209 $\\ 
Stacking           & $ 1259 $ & $ 7.94 $ & $ 309 $ & $ 3.236 $\\ 
\tableline
\end{tabular}
\end{center}
\end{table}
\clearpage 

Overall we find that the B-Splines, IDW and RBF methods reconstruct
the ellipticity and size most accurately (see Gentile et al., 2012),
with $\sigma(e)\approx 2.5$x$10^{-4}$ and
$\sigma(R^2)/R^2\approx 7.4$x$10^{-4}$ over all sets\footnote{B-Splines
  also achieved the highest leaderboard $P$ value.}. We note however
that this is a snapshot of performance and that
further investigations into tunable aspects of code 
could result in improvements in all methods. 

We summarise the behaviour of the submissions below. In each test all
other parameters are kept fixed except those discussed (with fiducial
values of $1000$ known star positions, no mask, and telescope
parameters given in Section \ref{Description of the Simulations}). We
refer to Figures \ref{fig00} and \ref{fig0} that show the change in
the inverse variance of
the reconstructed PSFs over the fiducial sets (set 1 for Airy, and set
8 for Moffat profiles, see Table \ref{set1}) 
when each of the categories is varied. In Figures \ref{fig1} to
\ref{fig5} we show pictographic tables of results. 

\begin{itemize}
\item 
{\bf PSF Type}.
For the best performing methods we find a trend that both ellipticity
and size are estimated more accurately for the Airy PSF than for the the
Moffat PSF. 
\item
{\bf Addition of Kolmogorov Power}. For each set combination where
both Moffat and Moffat-plus-Kolmogorov power are available (e.g. the 
4-arm $+$ masks) we find evidence for methods performing less well
with the addition of Kolmogorov power (see also Figures \ref{fig1},
\ref{fig2}, \ref{fig3}). In Figure \ref{fig4} we also show the impact
of adding a Kolmogorov power spectrum to a set that uses an
Airy PSF profile. We find that the addition of this random
component degrades the residual ellipticity reconstruction by a 
factor of $\gs 2-5$, but has less impact on size reconstruction, as
expected since the power is in ellipticity only. These 
results will necessarily depend on the amplitude of the assumed power
spectrum, this will vary for each ground-based telescope, and 
knowledge/information about this is improving 
(e.g. Heymans et al. 2012). 
In addition atmospheric turbulence also changes the PSF size, but we
do not simulate this here. It is possible that, depending on the site
and weather, the impact of turbulence may be weaker or stronger than
that simulated for this study.
\item
{\bf Masks}. We show results for the mask variation in Figure
\ref{fig1}. We find that for all methods the presence of diffraction
spikes does not  degrade the ability to measure the ellipticity of the
PSF. For the Airy function the diffraction
spikes act to increase the effective size of the PSF, this enables
methods to measure the fractional error $\sigma(R^2)/R^2$ more
accurately; but note that for a fixed $\sigma(R^2)$ a large size will
decrease the fractional error by definition. For the Moffat PSF
the diffraction spikes impact the size estimation significantly. 
We note however this was a simple addition of a mask
with no commensurate change in the variation of ellipticity or size
across a field of view, also the diffraction spikes contained low
flux (only observable with the eye if one stacked all stars) higher
signal-to-noise stars would change this, we leave an investigation of
these effects for future work.
\item
{\bf Number of Stars}. We find that all methods are only weakly dependent, or
insensitive to the number of stars used to reconstruct the PSF in these
simulations, except for those sets
in which we include a Kolmogorov power spectrum where we find that a
larger number of stars results in a better reconstruction for the best
methods (see Figure
\ref{fig2}). This indicates that PSFs with spatial power on smaller scales
require more stars for a particular reconstruction accuracy than PSFs
without power on small spatial scales. 
\item
{\bf Size of PSF} For the Airy profile we find that the larger
the PSF the more accurately its size can be measured, for the Moffat
we find a weak dependence with size. 
This is understandable because a larger PSF
is better sampled and hence the size is easier to measure. 
However we stress that an increase of the size of the PSF relative to
the apparent size of galaxies will cause the galaxies to be less
well-resolved, losing information and placing greater demands on
shape measurement \citep{sph08}. Also with the simulations presented
the impact of sampling on weak lensing shape measurement was not tested, 
only the performances of the PSF interpolation methods. 
We show results for the PSF size variation in Figure
\ref{fig3}. When trading requirements of PSF model residuals against requirements
for resolution (i.e. the absolute size of the ellipticity and PSF)
such behaviour should be noted.
\item 
{\bf Telescope Parameters}. We show results for the PSF size variation in Figure
\ref{fig5}. In varying the telescope parameters in the
\citet{jsj08} model we change the fiducial parameters respectively 
$(a_0= 0.014$, $d_0=-0.006$, $c_0=-0.010)$, to $a_0= -0.011$,
$d_0=0.009$ and $c_0=-0.011$ i.e. an opposite astigmatism, a positive
de-focus and a $10\%$ increased coma. We find that methods in this
experiment were not affected by the change in defocus, but performed
better with the change in these astigmatism and coma parameters.  
\end{itemize}

We discuss each method individually in Appendix A. 
\clearpage

\clearpage

%%%%%%%%%%%%%%%%%%%%%%%%%%%%%%%%%%%%%%%%%%%%%%%%%%%%%%%%%%%%%%%%%%%%%%
\section{Conclusions}
\label{Conclusions}

This paper presents the first blind simulation challenge aimed to
test optical PSF reconstruction methods. Simulations were generated
in which participants were presented with a spatially
varying PSF, sparsely sampled by stars, and asked to reconstruct the
PSF at non-star positions. The competition, the GREAT10 Star
Challenge, attracted 30 submissions from 9 teams; several of these
teams were from non-astronomy backgrounds. The simulation presented
participants with $27$,$500$ stars over 1300 images subdivided into 26
sets, where in each set a category change was made in the type or
spatial variation of the PSF. The simulations were intentionally simplistic, so
as to present the problem in an approachable way; in
particular the spatial variation of the PSF and the form of
the PSF use simple analytic functions. In addition only spatial
variation, not temporal variation, was tested; hence these results
should not be used to make specific statements about any particular
experiment but should provide a benchmark with which methods can be
tested and improved\footnote{Data is available for download here {\tt
    http://great.roe.ac.uk/data/solutions/}.}

In this paper we analyse the submissions by testing how well each one
can measure the ellipticity and size of the PSF. We quantify this as
the inverse variance in the modelled PSF in each image for ellipticity and
sized-squared -- defined using weighted quadrupole moments. This study
was motivated by a desire to find methods that will be of use for weak
gravitational lensing, where the PSF must be reconstructed to high
accuracy \citep{sph08,sparsity} at galaxy positions, but these results
should also
be of more general interest for any science case that analyses
galaxy images with optical data. 

The submissions, and this paper, present a snapshot of any methods'
ability to model the PSF. Due to the nature of the competitive
blind submissions post-challenge tuning of methods, that 
may yield significant improvements for any given
method over the results presented here (see Gentile et al., 2012 for example), were not
investigated. Each method submitted is summarised
in Appendix A. We can however make some general statements about  
regimes in which methods tend to perform well or poorly when run in a blind way.

The functional form of the PSF was either a Moffat function or a
Airy function, the spatial variation of the PSF
was modelled using the analytic function given in \citet{jsj08}, in
addition we optionally included diffraction spikes ($+$ or $\ast$
forms), changed the PSF size (from 3.0 pixels to 1.5 or 6.0 pixels),
changed the number of stars (from 1000 to 500 or 2000), and added an
atmospheric turbulence pattern in ellipticity (with a Kolmogorov power
spectrum). To summarise the conclusions we find that 
\begin{itemize}
\item 
The best methods can reconstruct the PSF with an accuracy of $\sigma(e)\approx 2.5$x$10^{-4}$ and
$\sigma(R^2)/R^2\approx 7.4$x$10^{-4}$ over all sets. 
\item 
Methods that performed poorly did so in part because the functional
form of the PSF was not modelled correctly (in particular the
Airy function).
\item 
Smaller PSFs were more difficult to model than larger PSFs for the
Airy function. But we add a caution that this does not mean
larger PSFs are better for weak lensing, because information on a
target object is lost; instead this means that well sampled PSFs are
better for weak lensing. 
\item 
Diffraction spikes caused the size of Moffat PSFs to be modelled less
accurately, but Airy PSFs more accurately, due to the increase
in the effective size.
\item 
The addition of atmospheric Kolmogorov power (equivalent to short
exposure PSFs, see Heymans et al., 2012) made ellipticity and size
reconstruction less accurate by a factor of $\gs 2-5$ for all methods. 
We add the caveat that the temporal nature of varying PSFs was not
investigated, therefore
methods such as cross-correlation between sequential images, that
could potentially improve modelling, were not investigated.
\end{itemize} 
For subsequent blind PSF modelling challenges the
realism of the temporal and wavelength dependent nature 
of PSF variation could be included,
and the simulations could be tailored to specific experiments.  

Modelling the PSF is of critical importance in efforts to understand
the nature of dark energy and dark matter using weak gravitational
lensing; where any inaccuracy in the modelled PSF can cause biases, and
increased errors of cosmological parameters of interest. To address
this crucial open problem this initial
presentation of a blind PSF reconstruction challenge will hopefully provide a
benchmark upon which methods can continue to be refined and tested.

\newpage
\acknowledgements{\small {\em Acknowledgements:} TDK is supported by a Royal Society
University Research Fellowship, and was supported by an Royal
Astronomical Society 2010 Fellowship for some of this
work. BR and CH acknowledge support from the the European Research Council
in the form of a Starting Grant with numbers 24067 (BR) and 240185 (CH). 
RM is supported by a Royal Society University Research
Fellowship. 
DG was supported by SFB-Transregio 33 ‘The Dark Universe’ by the
Deutsche Forschungsgemeinschaft (DFG) and the DFG cluster of
excellence `Origin and Structure of the Universe' and thanks Gary
Bernstein and Stella Seitz for helpful discussions. MGe, GC, GM are
supported by the Swiss National Science Foundation (SNSF). GL thanks 
Wei Cui for useful discussions. GL and BX were supported in part by the
U.S. Department of Energy through Grant DE-FG02-91ER4068 and GL is also
supported by the one-hundred talents program of the Chinese Academy of
Sciences (CAS). MK thanks Liping Fu.
This work was funded by a EU FP7 PASCAL 2 Challenge
Grant. Workshops for the GREAT10 challenge were funded by the eScience
STFC Theme and the by NASA JPL, and hosted at the eScience Institute
Edinburgh and by IPAC Caltech Pasadena. We thank Francesca
Ziolkowska, Harry Teplitz and Helene Seibly for local organization of
the workshops. We thank the GREAT10 Advisory team, co-authors of
GREAT10 handbook \citep{great10}, for discussions before and after the challenge.}

{\small {\em Contributions:} All authors contributed to the writing
  and analysis presented. TDK was PI of GREAT10, defined and created the
  simulations, and lead the analysis. 
  TDK, BR, MG, CH, RM were active members of the GREAT10 team during (12/2010 to 09/2011)
  and after the challenge. BR created the FITS image simulation code.   
  MGe, FC, GM, DG, MK, KG, AS, AM, GL, BX submitted entries to the GREAT10 star challenge. 
  DW maintained the GREAT10 leaderboard and processed submissions with
  TDK during the challenge.}
%__________________________________________________________________

%_________________________________________________________________

\newpage
\section*{Appendix A: Description of Methods}
Here we include a brief description and references for each of the
methods submitted to the challenge. 

Several methods use the name ``Kriging'', which is
in fact the same method as Gaussian process regression, the method
submitted for the methods MoffatGP; Kriging is a different term which has been in use in
the geostatistics field but all are types of Gaussian processes.

\subsection*{PSFEx (Gruen)}
PSFEx uses version 3.9.1 of the PSFEx software
\citep{PSFEx}. The method models the PSF using a
functional basis, the coefficients of which are allowed to vary with a
polynomial dependence on the position in the field. Details of the
configuration can be found in the PSFEx manual\footnote{\tt
  http://www.astromatic.net/software/psfex}. For the GREAT10
submission, the functional basis is chosen to be a sub-pixel grid,
from which PSF images on the input pixel scale are produced using
Lanczos interpolation of order 4. In order to improve configuration
parameters, the P metric is calculated on stars reserved from the fit.
For this a Gaussian weight function with much smaller scale than in
the final analysis (3pix FWHM) is used in order to suppress the noise
in the images. The spatial variation was chosen to be of order 8 (4) on
the sets with (without) atmospheric Kolmogorov power and the size of
the sub-pixel grid to be 1/4.7 of the PSF FWHM on all sets except the
very undersampled sets 6, 14 and 26, where a scale of 0.25pix is used
instead. Note that these choices were made without knowledge of the
true properties of the sets. 

\subsection*{PCA+Kriging (Li, Xin)}

The basic idea is to find the principal components of ensemble of stars in an image.
To find the right principal components (PCs), were needed to align all of the stars at a same
center. Here a fast algorithm \citep{li12} was used to locate the
centroid for each star and then ordinary Kriging fitting algorithm was
used to reproduce the star, whose center was exactly located at the
center of stamp grid. Each star was represented by 5 PCs, with five
corresponding coefficients. According to the noise in the star stamp,
an additional Gaussian noise component was included in each pixel and re-evaluated the
corresponding coefficients in ten realisations; this helped us to estimate the
uncertainties for each coefficient of PCs and each star. Ordinary
Kriging fitting process was the used to predict the value of each
coefficient at the asked positions and the new stars were composed of
these fives PCs with the predicted coefficients. 

\subsection*{Gaussianlets (Li, Xin)}

Gaussianlets is a simplified version of shapelets without any angular
components, i.e., there are no shapelets with $m\not =0$. The
ellipticity of each star was calculated at the first step, then the
size of each star $R_i$ was estimated quickly using the algorithm
described in \citet{li12}.  One set of unique gaussianlets with a
maximum order $n_{\rm max}=4$ were created with
$R=\langle R_i\rangle$. 
These gaussianlets are circularly symmetric and were then reshaped into elliptical
profiles according to the ellipticities that were measured in the first step to
fit an individual star. The coefficients of gaussianlets were
calculated by minimizing a chi-square function. Finally each star
was described by $7$ 
parameters, $e_1$, $e_2$ and the five coefficients of gaussianlets. Ordinary Kriging
interpolation was then used to predict these seven parameters at the asked
positions. To reproduce the expected virtual star, the gaussianlets
were reshaped according to $e_1$ and $e_2$ and were added up together according
to their coefficients. 

\subsection*{B-Splines (Gentile, Courbin, Meylan)}

The B-Splines method, like the IDW, RBF and Kriging schemes also
described in this article, uses the same underlying PSF estimation
scheme that consists of the following stages.  
First, an elliptical Moffat profile is fitted to each star at known
position. Fitting is performed using a custom-developed minimizer
based on an ``adaptive cyclic coordinate descent algorithm''. This
minimizer is also used in the \emph{gfit} galaxy shape measurement
method described in \citet{great10}. Second, an analysis is
performed of the spatial distribution of each Moffat parameter across
the image. Third, a spatial interpolation scheme is adopted (here
B-Splines) to predict the values of each Moffat parameter $p$ at asked
positions. Finally, pixelized star images are reconstructed at asked
positions, based on the interpolated Moffat profile. 

B-Splines perform a spatial interpolation of individual Moffat PSF
parameters using the bivariate basis-spline algorithm described in
\citet{Dierckx80,Dierckx95} and implemented in the Python SciPy
\mbox{interpolate} module. The main parameters affecting the
interpolation are the degree of the spline, the number of knots, and a
smoothing factor. A 3rd order spline was used but the algorithm was
allowed to automatically optimize the number of knots and the
smoothing factor. 
A more thorough description of B-Splines, as well as the IDW, RBF and
Kriging interpolation methods can be found in \citet{Gentile12}.   

\subsection*{Inverse Distance Weighting (IDW) (Gentile, Courbin, Meylan)}

The IDW interpolation algorithm \citep{Shepard68} is used to
interpolate the Moffat parameters of the fitted PSF (see
B-splines). Weights are allocated to the stars or
parameters to interpolate. The closer the observations from a target
location, the greater the weight ascribed to them. The estimated value
of the parameter at the target point is a weighted sum of the values
of all neighboring observations considered. The weighting power
$\gamma$ determines how fast the weights tend to zero as distances
increase. The Star Challenge results were obtained with $\gamma=2$
with a neighborhood size between $5$ and $15$ pixels depending on the
density of stars. 

\subsection*{Radial Basis Function (RBF and RBF-thin) (Gentile, Courbin, Meylan)}

The RBF and RBF-thin methods make use of Radial Basis Functions to
predict the values of the PSF parameters at non-star positions. As in
B-splines the PSF is approximated by a Moffat profile. A
Radial Basis Function \citep{Buhmann03, Press07}, is a
radially-symmetric, real-valued function, whose value at a target
location only depends on the distance to some other point. The
prediction at a target location is based on the weighted sum of the
RBFs evaluated in a neighborhood centered at that location. 

The RBF and RBF-thin methods respectively use the linear and thin-plate
functions. Their implementation is based on the interpolation function
available in the Python SciPy \mbox{interpolate} module with a
neighborhood size between $25$ and $30$ pixels. For the submission to
the challenge smoothing was disabled, i.e., exact interpolation was
used where the PSF reconstructed at known positions should be exactly the
input data. 

\subsection*{Kriging (Gentile, Courbin, Meylan)}

Ordinary Kriging \citep[e.g.][]{Waller04, Webster07} is used
to interpolate PSF parameters (Moffat profiles as in
B-splines) across the PSF field. For the Star Challenge,
a unique implementation was created in Python for greater flexibility
and control of the algorithm. In this version, no attempt was made to
correct for any spatial anisotropy or drift found in the data. 
The experimental variograms were fitted using the Levenberg-Marquardt
\citep{Levenberg44, Marquardt63} fitting function from the SciPy
optimize module. The program dynamically selects the theoretical
variogram models and parameters that produce the best fit. The area
used for interpolation is a circular area with a radius between $700$
and $1000$ pixels from the center of the $4800\times4800$ PSF
field. Lag distances were selected in the range $100 \leq h \leq 300$
pixels depending on the image and the PSF model parameter to
estimate. The number of observations $N$ to include in the
interpolation neighborhood was typically $5 \leq N \leq 20$ depending
on the image star density. As a rule of thumb, a tolerance was adopted
for the distances $\Delta h \thickapprox h/2$ and angles $\Delta \theta = 22.5
^\circ$.  

\subsection*{IDWStk (Gentile, Courbin, Meylan)}

The IDWStk method experimented with an algorithm whereby the star postage
stamp to reconstruct at asked position is estimated by \emph{stacking
  the pixels} of nearby, surrounding stamps located at known
positions. Each pixel carries a weight that depends on its distance to
the location where reconstruction has to take place. These weighting
factors are calculated using Inverse Distance Weighting (IDW). For the
Star challenge, the number of surrounding nearby stars in the stacking
was typically $10$. 

\subsection*{MoffatGP (Georgatzis, Mariglis, Storkey)}
First, each 48x48 star image was reduced to a 30x30 image. 
Predictions for the coefficients were made at the asked positions (with
their corresponding offsets) and then the star images at test positions
were reconstructed using the Moffat function (generating a 48x48 image for
each star patch). Five coefficients 
per star patch were produced and then used as training outputs for the regression
method. Regression was performed using the Gaussian Process (GP) framework
on an augmented input space. Along with the stars' centre locations, for each
star patch a distinct variable the offset of the star centre from
the bottom left corner of the centre pixel was isolated, and provided as an additional input
to the GP. The neural network covariance function \citet{Rasmussen06} was chosen to encode
correlations between data points. Predictions for the coefficients were made at
the asked positions (with their corresponding offsets) and then the star images
at test positions were reconstructed using the Moffat model (generating a 48x48
image for each star patch). The method is described in more detail in \citet{Georgatzis11}.


\begin{thebibliography}{}

\bibitem[\protect\citeauthoryear{Albrecht et al.}{2001}]{arev} Albrecht A.\ et al., 2001
\bibitem[\protect\citeauthoryear{Bacon et al.}{2003}]{bac03} Bacon D.\ et al., 2003, MNRAS 344, 673
\bibitem[\protect\citeauthoryear{Bartelmann \& Schneider}{2001}]{bs01} Bartelmann M.\ \& Schneider P., 2001, Phys.\ Rep.\ 340, 291
\bibitem[\protect\citeauthoryear{Bernstein \& Jarvis}{2002}]{bj02}
  Bernstein G.\& Jarvis M., 2002, AJ, 123, 583
\bibitem[\protect\citeauthoryear{Bertin}{2011}]{PSFEx} Bertin, E.\ 2011, Astronomical 
Data Analysis Software and Systems XX, 442, 435 
\bibitem[\protect\citeauthoryear{Bridle et al.}{2010}]{great08} Bridle
  S.\ et al., 2010, MNRAS 405, 2044
\bibitem[\protect\citeauthoryear{Buhmann}{2003}]{Buhmann03}
  Buhmann, M. D. 2003, Radial basis functions: theory and
  implementations, Vol. 12 (Cambridge University Press), 274
\bibitem[\protect\citeauthoryear{Chang et al.}{2012}]{lsstpsf} Chang C.\ et al., 2012, MNRAS submitted (arXiv:1206.1378)
\bibitem[\protect\citeauthoryear{Cropper et al.}{2012}]{chk12} Cropper M., Hoekstra H., Kitching T.\ et al., 2012, MNRAS submitted
\bibitem[\protect\citeauthoryear{Cypriano et al.}{2010}]{cyp09}
  Cypriano E., Amara A., Voigt L., Bridle S., Abdalla F.,
  R\'efr\'egier A., Seiffert M.\ \& Rhodes J., 2010, MNRAS 405, 494
\bibitem[\protect\citeauthoryear{Dierckx}{1980}]{Dierckx80}
  Dierckx, P. 1980, An algorithm for surface fitting with spline
  functions (Katholieke Univ. Leuven)
\bibitem[\protect\citeauthoryear{Dierckx}{1995}]{Dierckx95}
  Dierckx, P. 1995, Curve and surface fitting with splines, Monographs
  on numerical analysis (Clarendon Press)
\bibitem[\protect\citeauthoryear{Heymans et al.}{2006}]{step1} Heymans C.\ et al., 2006, MNRAS 368, 1323
\bibitem[\protect\citeauthoryear{Heymans et al.}{2012}]{cfhtpsf} Heymans C.\ et al., 2012, MNRAS 421, 381
\bibitem[\protect\citeauthoryear{Hoekstra}{2004}]{henkpsf} Hoekstra H., 2004, MNRAS 347, 1337
\bibitem[\protect\citeauthoryear{Hoekstra, Yee \& Gladders}{2004}]{hyg04} Hoekstra H., Yee H.\ \& Gladders M., 2004, ApJ, 606, 67
\bibitem[\protect\citeauthoryear{Hoekstra \& Jain}{2008}]{hrev}Hoekstra H.\ \& Jain B., 2008, Ann.\ Rev.\ Nuc.\ Part.\ 58, 99
\bibitem[\protect\citeauthoryear{Hu}{1999}]{hu99} Hu, W.\ 1999, \apjl,
  522, L21 
\bibitem[\protect\citeauthoryear{Gentile et al.}{2012}]{Gentile12}
  Gentile, M., Courbin, F., \& Meylan, G. in prep, A\&A
\bibitem[\protect\citeauthoryear{Georgatzis}{2011}]{Georgatzis11}Konstantinos
  Georgatzis. Machine learning methods for regression in as-
  tronomical imaging. Master's thesis, University of Edinburgh, School
  of Informatics, 2011.
\bibitem[\protect\citeauthoryear{Jarvis \& Jain}{2005}]{jj05} Jarvis
  M.\ \& Jain B., 2005, ApJ submitted (astro-ph/0412234)
\bibitem[\protect\citeauthoryear{Jarvis et al.}{2008}]{jsj08} Jarvis, M., Schechter, P., \& Jain, B.\ 2008, arXiv:0810.0027
\bibitem[\protect\citeauthoryear{Jain, Jarvis \& Bernstein}{2006}]{jain06} Jain B., Jarvis M.\ \& Bernstein G., 2006, JCAP 0602, 001
\bibitem[\protect\citeauthoryear{Kitching et al.}{2012a}]{great10}
  Kitching T.\ et al., 2012a, AoAS 5, 2231
\bibitem[\protect\citeauthoryear{Kitching et al.}{2012b}]{g10res} Kitching T.\ et al.\ 2012b, MNRAS 423, 3236
\bibitem[\protect\citeauthoryear{Kitching et al.}{2012c}]{kaggle}
  Kitching T.\ et al.\ 2012c, New Astronomy Reviews
\bibitem[\protect\citeauthoryear{Kuijken}{2006}]{kk06} Kuijken,
  K.\ 2006, arXiv:astro-ph/0610606 
\bibitem[\protect\citeauthoryear{Levenberg}{1980}]{Levenberg44}
  Levenberg, K. 1944, The Quarterly of Applied Mathematics, 2, 164 
\bibitem[\protect\citeauthoryear{Li et al.}{2012}]{li12} Li, G., Xin, B., \& Cui, W.\ 2012, arXiv:1203.0571 
\bibitem[\protect\citeauthoryear{Massey et al.}{2007}]{step2} Massey R.\ et al., 2007b, MNRAS 376, 13
\bibitem[\protect\citeauthoryear{Massey, Kitching \& Richard}{2010}]{mrev} Massey R., Kitching T.\ \& Richard J., 2010, Rep.\ Prog.\ Phys.\ 73, 086901
\bibitem[\protect\citeauthoryear{Massey et al.}{2012}]{mhk12} Massey
  R., Hoekstra H., Kitching T.\ et al., 2012, MNRAS submitted
\bibitem[\protect\citeauthoryear{Marquardt}{1963}]{Marquardt63}
  Marquardt, D. W. 1963, Journal of the Society for Industrial and
  Applied Mathematics, 11, 431
\bibitem[\protect\citeauthoryear{Moffat}{1969}]{moffat} Moffat,
  A.~F.~J.\ 1969, \aap, 3, 455 
\bibitem[\protect\citeauthoryear{Papageorgiou}{2011}]{Papageorgiou11}Asimakis
  M. Papageorgiou. Point spread function modelling in astronomical
  imaging. Master's thesis, University of Edinburgh, School of
  Informatics, 2011.
\bibitem[\protect\citeauthoryear{Paulin-Henriksson et al.}{2008}]{sph08} Paulin-Henriksson S., Amara A., Voigt L., R\'efr\'egier A.\ \& Bridle S., 2008, A\&A 484, 67
\bibitem[\protect\citeauthoryear{Paulin-Henriksson et al.}{2009}]{sparsity} Paulin-Henriksson S., R\'efr\'egier A.\ \&
  Amara A., 2009, A\&A 500, 647
\bibitem[\protect\citeauthoryear{Plazas \& Bernstein}{2012}]{plaz}
  Plazas, A.~A., \& Bernstein, G.~M.\ 2012, arXiv:1204.1346 
\bibitem[\protect\citeauthoryear{Press et al.}{2007}]{Press07} Press,
  W. H., Teukolsky, S. A., Vetterling, W. T., \& Flannery, B. P. 2007, 
Numerical Recipes 3rd Edition: The Art of Scientific Computing, 3rd edn.
(Cambridge University Press)
\bibitem[\protect\citeauthoryear{Rasmussen et al.}{2006}]{Rasmussen06}Rasmussen C.E. and Williams
  C.K.I. Gaussian Processes for Machine Learning. Adaptive Computation and Machine Learning. MIT Press, Cam-
bridge, MA, USA, 2006.
\bibitem[\protect\citeauthoryear{R\'efr\'egier}{2003}]{rrev}
  R\'efr\'egier A., 2003, ARA\&A 41, 645
\bibitem[\protect\citeauthoryear{Rowe}{2010}]{rowe10} Rowe, B.\ 2010, \mnras, 404, 350 
\bibitem[\protect\citeauthoryear{Rhodes et al.}{2007}]{rho07} Rhodes J.\ et al., 2007, ApJS 172, 203
\bibitem[\protect\citeauthoryear{Schrabback et al.}{2010}]{sch10}
  Schrabback T., Hartlap J., Joachimi B.\ et al., 2010, A\&A 516, 63
\bibitem[\protect\citeauthoryear{Shepard}{2007}]{Shepard68} Shepard,
  D. 1968, in ACM ’68: Proceedings of the 1968 23rd ACM national conference (New York, NY, USA: ACM), 517–524 
\bibitem[\protect\citeauthoryear{van Waerbeke et al.}{2005}]{lvw05} van Waerbeke, L., Mellier, Y.\ \& Hoekstra, H. 2005, A\&A, 429, 75
\bibitem[\protect\citeauthoryear{Voigt et al.}{2011}]{voi11} Voigt L.,
  Bridle S., Amara A., Cropper M., Kitching T., Massey R., Rhodes
  J.\ \& Schrabback T. 2011, MNRAS submitted, arXiv:1105.5595
\bibitem[\protect\citeauthoryear{Waller}{2004}]{Waller04} Waller, L. \&
  Gotway, C. 2004, Applied spatial statistics for public health data,
  Wiley series in probability and statistics (John Wiley \& Sons) 
\bibitem[\protect\citeauthoryear{ Webster \& Oliver}{2007}]{Webster07}
  Webster, R. \& Oliver, M. 2007, Geostatistics for environmental
  scientists, Statistics in practice (Wiley)
\bibitem[\protect\citeauthoryear{Weinberg et al.}{2012}]{wrev}
  Weinberg D.\ et al., 2012
\bibitem[\protect\citeauthoryear{Wells et al.}{1981}]{fits} Wells,
  D.~C., Greisen, E.~W., \& Harten, R.~H.\ 1981, \aaps, 44, 363 






\end{thebibliography}
\end{document}